\documentclass[11pt,a4paper]{article}
\usepackage[utf8]{inputenc}
\usepackage[english]{babel}
\usepackage[a4paper, left=1.93cm, right=1.93cm, top=2.54cm, bottom=4.94cm]{geometry}
\usepackage{amsmath, amssymb}
\usepackage{graphicx}
\usepackage{float}
\usepackage{caption}
\usepackage{hyperref}
\usepackage{titlesec}
\usepackage{enumitem}
\usepackage{fancyhdr}
\usepackage[super]{natbib}
\fancypagestyle{plain}{
  \fancyhf{} 

}

\usepackage[font={it}]{caption}
\DeclareCaptionFont{tenpt}{\fontsize{10pt}{12pt}\selectfont}
\titleformat{\section}
  {\centering\normalfont\bfseries\fontsize{12pt}{14pt}\selectfont}
  {\thesection}{1em}{}

\titleformat{\subsection}
  {\normalfont\bfseries\fontsize{11pt}{13pt}\selectfont}
  {\thesubsection}{1em}{}

\usepackage{helvet}

\usepackage[T1]{fontenc}
\title{\fontsize{16pt}{6pt}\selectfont\bfseries External Validation of Deep Learning Models for BI-RADS Breast Density Prediction from Ultrasound Images}

\author{
    Yuxuan Chen$^{a}$, Arianna Bunnell$^{c,d}$, Yanqi Xu$^{e}$, Haoyan Yang$^{f}$, Thomas K. Wolfgruber$^{c}$, \\
    John A. Shepherd$^{c}$, and Yiqiu Shen$^{b,*}$ \\[1ex]
    \normalsize $^{a}$Perlmutter Cancer Center, NYU Langone Health, New York, NY, USA \\
    \normalsize $^{b}$Department of Radiology, NYU Langone Health, New York, NY, USA \\
    \normalsize $^{c}$University of Hawai’i Cancer Center, Honolulu, HI, USA \\
    \normalsize $^{d}$University of Hawai’i at Mānoa, Honolulu, HI, USA \\
    \normalsize $^{e}$Center for Data Science, New York University, New York, NY, USA \\
    \normalsize $^{f}$Department of Computer Science, Stony Brook University, Stony Brook, NY, USA \\ [0.5ex]
    \normalsize $^{*}$Corresponding Author: \texttt{Yiqiu.Shen@nyulangone.org}
}
\begin{document}
\pagestyle{empty}
\thispagestyle{empty}
\date{} 
\maketitle

\begin{center}
\textbf{Abstract} \\ [0.5ex]
\end{center}

\textbf{Purpose:} To externally validate three deep learning models (DenseNet121, ViT-B/32, and ResNet50) for predicting mammographic breast density from breast ultrasound exams on an independent cohort. \\ [1ex]
\textbf{Methods:} The external validation set comprised 2,000 ultrasound exams, including 500 cancer cases defined by an initial negative exam (BI-RADS 1 or 2) followed by a cancer diagnosis within 6 months to 10 years, and 1,500 negative controls matched by manufacturer and study year. Performance was measured using patient-level AUROC in four density categories: A (fatty), B (scattered), C (heterogeneous), and D (extremely dense). In addition, we performed risk prediction using AI-derived density as a downstream assessment. We evaluated 10-year risk prediction by incorporating age and AI-derived density into the Tyrer–Cuzick model and comparing risk prediction performance with a reference model using age and mammography-reported density. \\ [1ex]
\textbf{Results:}
In the external validation, all three models performed best in extremely dense breasts (0.868--0.899), with strong performance in fatty (0.814--0.838) and scattered density (0.764--0.799), and lower performance in heterogeneously dense breasts (0.699--0.729). DenseNet121 achieved the highest overall performance (micro-averaged AUROC 0.885). Performance across different categories was comparable between internal and external testing. In the risk modeling, age combined with AI-derived density yielded lower AUROC than age combined with mammography-reported density (0.541 vs. 0.570; p = 0.23), with no statistically significant difference.\\ [1ex]
\textbf{Conclusions:} Deep learning models generalize well to external data with different racial composition for breast density assessment. While performance is strongest in extremely dense breasts, heterogeneously dense remains more challenging, highlighting the need for targeted optimization.

\section{INTRODUCTION}

Mammographic breast density is one of the strongest independent risk factors for breast cancer and is incorporated into multiple clinical risk prediction models \cite{tice2008using}. It is most commonly defined using the BI-RADS categorization system \cite{spak2017bi}. However, breast density is derived from mammography, which is not universally available, particularly in low-resource or ultrasound-based screening settings \cite{sood2019ultrasound}. Recent work demonstrated that deep learning models can accurately estimate mammographic BI-RADS breast density directly from clinical breast ultrasound images and achieve performance comparable to mammographic density when used in modeling breast cancer risk \cite{bunnell2025prediction}. The purpose of this study is to validate these ultrasound-based models to predict mammographic breast density in an external cohort with a different demographic composition, assessing model generalizability and comparing performance with the original results on the internal test set.

\section{METHODS}

\subsection{External Validation Cohort}
We constructed an external validation dataset from NYU Langone Health System that includes screening and diagnostic breast ultrasound examinations. The dataset included 500 cancer cases and 1,500 negative controls. Cancer cases were defined as patients with an initial negative ultrasound examination (BI-RADS 1 or 2) who subsequently developed pathology-confirmed breast cancer between 6 months and 10 years after the index exam. The controls consisted of patients with negative ultrasound (BI-RADS 1 or 2), no pathologically confirmed malignancy, and at least 2 years of cancer-free follow-up. Study year and ultrasound manufacturer were strictly matched between cases and controls. We followed the established protocol \cite{Shamout2021TheNB} for extracting ground truth breast density from patient’s prior mammography reports. All ultrasound images were preprocessed using an established pipeline \cite{bunnell2024busclean} to remove operator-dependent artifacts, including Doppler overlays, text annotations, and caliper markers, to reduce bias.

\subsection{Breast Density Inference and Risk Modeling}
 We externally validated three deep learning models (DenseNet121, ViT-B/32, ResNet50) trained on a large multi-institutional dataset from Hawaii comprising over 302,574 images from 10,393 women collected between 2009 and 2022~\cite{bunnell2025prediction}. On average, each ultrasound exam contains 16.2 images (SD: 9.8). Each model outputs a four-dimensional softmax probability vector per image, corresponding to BI-RADS categories (A: fatty, B: scattered, C: heterogeneous, D: extremely dense). Image-level probabilities were averaged across all images within an exam to obtain a patient-level probability vector, and the category with the maximum averaged probability was assigned as the predicted density. Following the risk analysis from the original study, AI-derived breast density was incorporated into the Tyrer–Cuzick (TC) model~\cite{tyrer2004breast} for 10-year risk prediction. Risk assessments were conducted using TC model version 8, with all TC scores computed using the IBIS Breast Cancer Risk Evaluation Tool \cite{cuzick2017ibis}. Performance was compared between TC models parameterized with age and AI-predicted breast density versus age and ground-truth breast density.

\subsection{Statistical Analysis}
Model performance was evaluated using the Area Under the Receiver Operating Characteristic Curve (AUROC). For each BI-RADS density category, a one-vs-rest AUROC was computed, where positive cases were defined as patients whose ground-truth density matched the target category and negative cases as those belonging to the other three categories. The prediction score was the patient-level softmax probability assigned to the target category. Overall multiclass performance was summarized using the micro-averaged AUROC. For 10-year breast cancer risk prediction, AUROC was computed using the TC risk score as the prediction score, with cancer cases as positives and matched controls as negatives. Statistical uncertainty was quantified using 1,000 bootstrap resamples to derive 95\% confidence intervals (CIs), and pairwise comparisons of AUROC were performed using the two-sided DeLong test, with statistical significance defined as p < 0.05.

\begin{table}[H]
\footnotesize
\captionsetup{justification=raggedright, singlelinecheck=false}
\caption{\label{tab:demographics} Characteristics of the External Validation Dataset.}
\centering
\begin{tabular}{lccc}
\hline
\textbf{Characteristic} 
& \textbf{Overall (N = 2,000)} 
& \textbf{Cases (N = 500)} 
& \textbf{Controls (N = 1,500)} \\
\hline
\textbf{Patient Demographics} & & & \\
Mean Age (SD) & 59.9 $\pm$ 11.1 & 60.6 $\pm$ 11.0 & 59.7 $\pm$ 11.1 \\
\hline
\textbf{Ground Truth Density} & & & \\
A: Fatty & 61 (3.1\%) & 8 (1.6\%) & 53 (3.5\%) \\
B: Scattered & 610 (30.5\%) & 114 (22.8\%) & 496 (33.1\%) \\
C: Heterogeneous & 1,151 (57.6\%) & 330 (66.0\%) & 821 (54.7\%) \\
D: Extremely Dense & 178 (8.9\%) & 48 (9.6\%) & 130 (8.7\%) \\
\hline
\textbf{Study Year} & & & \\
Range & 2012--2019 & 2012--2019 & 2012--2019 \\
Median (IQR) & 2017 (2016--2017)  & 2017 (2016--2017)  & 2017 (2016--2017) \\
\hline
\textbf{Manufacturer} & & & \\
Siemens & 1,228 (61.4\%) & 307 (61.4\%) & 921 (61.4\%) \\
Philips & 772 (38.6\%) & 193 (38.6\%) & 579 (38.6\%) \\
\hline
\textbf{Race} & & & \\
White & 1,367 (68.4\%) & 370 (74.0\%) & 997 (66.5\%) \\
African American & 179 (9.0\%) & 40 (8.0\%) & 139 (9.3\%) \\
Asian & 80 (4.0\%) & 22 (4.4\%) & 58 (3.9\%) \\
Hispanic & 38 (1.9\%) & 5 (1.0\%) & 33 (2.2\%) \\
Unknown & 336 (16.8\%) & 63 (12.6\%) & 273 (18.2\%) \\
\hline
\end{tabular}
\end{table}

\section{RESULTS AND DISCUSSION}

\subsection{Dataset Characteristics}
Dataset characteristics are summarized in Table~\ref{tab:demographics}. The case and control cohorts were well balanced with respect to age (cases: 60.6 $\pm$ 11.0 years; controls: 59.7 $\pm$ 11.1 years). Ultrasound manufacturer distributions were identical by design (61.4\% Siemens, 38.6\% Philips), and examinations were matched by study year. Most examinations were heterogeneously dense (57.6\%) or scattered fibroglandular density (30.5\%), with fewer extremely dense (8.9\%) and almost entirely fatty (3.1\%) cases. The external population was predominantly White (68.4\%), differing from the population in Hawaii, where the original study was conducted and Asians are estimated to comprise approximately 60\% of the population \cite{bunnell2025prediction, DBEDT2024_SC_EST2023_SR11H_15}.

\subsection{External Validation Performance}
\begin{figure}[H]
\centering
\includegraphics[width=0.9\linewidth]{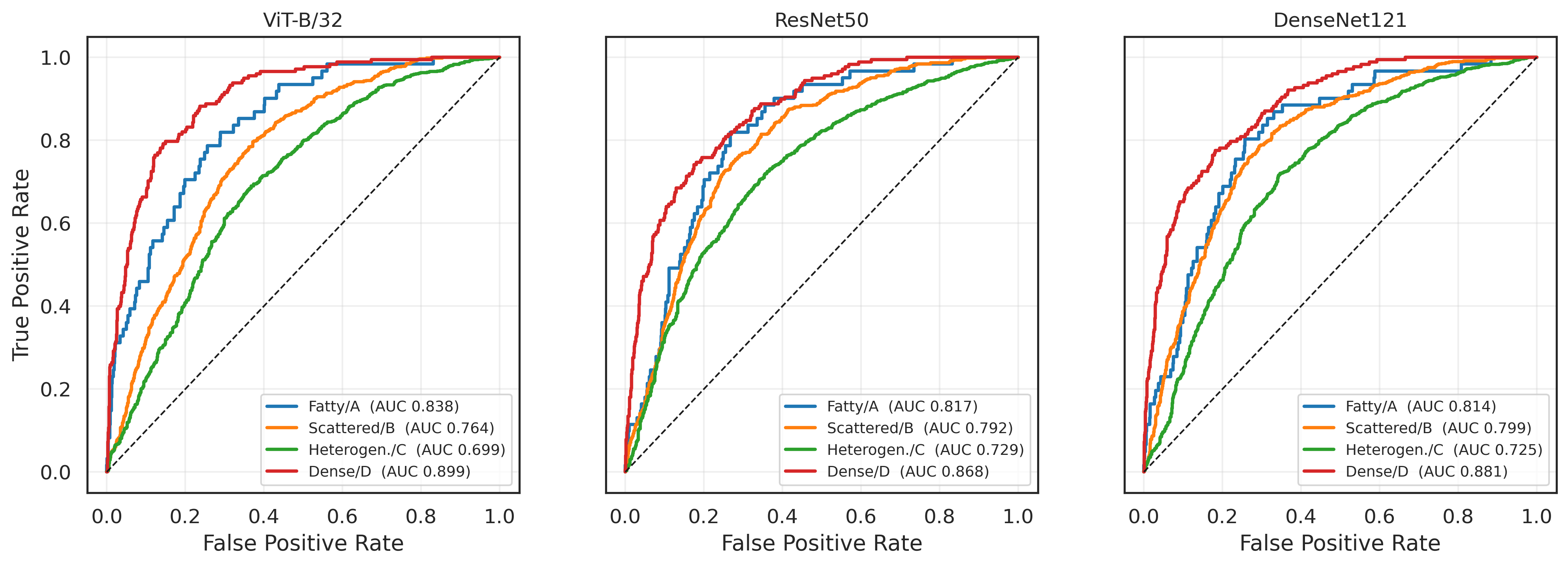}
\caption{\label{fig:frog}Receiver Operating Characteristic Curves from External Validation.}
\end{figure}

The external validation results are summarized in Table~\ref{tab:auroc}, with Receiver Operating Characteristic curves shown in Figure~\ref{fig:frog}. DenseNet121 achieved the highest overall performance (micro-averaged AUROC 0.885, 95\% CI: 0.875--0.894). Across models, performance was highest in extremely dense breasts (Category D; 0.868--0.899), remained strong in fatty (Category A; 0.814--0.838) and scattered fibroglandular density (Category B; 0.764--0.799), and was lower in heterogeneously dense breasts (Category C; 0.699--0.729). The same category-wise pattern was observed in internal testing~\cite{bunnell2025prediction}, where DenseNet121 achieved a micro-averaged AUROC of 0.854 (95\% CI: 0.842--0.866), with category-wise AUROCs of 0.879 (D), 0.841 (A), 0.787 (B), and 0.719 (C). Overall, external performance was comparable to internal testing, indicating good generalizability.

\begin{table}[H]
\footnotesize
\captionsetup{justification=raggedright, singlelinecheck=false}
\caption{\label{tab:auroc} External Validation Performance for Breast Density Prediction (95\% CI).}
\centering
\resizebox{\textwidth}{!}{
\begin{tabular}{lccccc}
\hline
\textbf{Model} & \textbf{Micro-Avg} & \textbf{Fatty/A} & \textbf{Scattered/B} & \textbf{Heterogen./C} & \textbf{Dense/D} \\
\hline
ViT-B/32 & 0.876 (0.867--0.885) & 0.838 (0.792--0.883) & 0.764 (0.743--0.785) & 0.699 (0.675--0.722) & 0.899 (0.878--0.920) \\
ResNet50 & 0.884 (0.876--0.894) & 0.817 (0.766--0.861) & 0.792 (0.773--0.811) & 0.729 (0.708--0.752) & 0.868 (0.844--0.893) \\
DenseNet121 & 0.885 (0.875--0.894) & 0.814 (0.760--0.858) & 0.799 (0.780--0.819) & 0.725 (0.700--0.748) & 0.881 (0.860--0.904) \\
\hline
\end{tabular}%
}
\end{table}

In 10-year risk modeling on the external cohort, the TC model combining age with AI-predicted density achieved an AUROC of 0.541 (95\% CI: 0.514--0.572), compared with 0.570 (95\% CI: 0.540--0.602) for the model using age and ground-truth density (p = 0.23). This performance difference was larger than that reported in the internal test set, where AI-derived and ground-truth density produced closely comparable 5-year risk prediction (AUROC 0.606 [95\% CI: 0.559--0.653] vs.\ 0.599 [95\% CI: 0.552--0.647]; p = 0.67). Nevertheless, across both cohorts, none of these differences in performance reached statistical significance.

\section{CONCLUSIONS}

Deep learning models generalize well to an external cohort with different racial composition for breast density assessment, with strongest performance observed in extremely dense breasts and reduced performance in heterogeneously dense breasts. While these results support broader clinical adoption, they also highlight the need for targeted optimization to improve performance in heterogeneously dense categories.


\bibliographystyle{unsrtnat}
\renewcommand{\refname}{\centering\fontsize{12pt}{14pt}\selectfont\bfseries REFERENCES}

\bibliography{references_sample}

\end{document}